\documentclass{article}

\usepackage{PRIMEarxiv}

\usepackage[utf8]{inputenc} 
\usepackage[T1]{fontenc}    
\usepackage{hyperref}       
\usepackage{url}            
\usepackage{booktabs}       
\usepackage{amsfonts}       
\usepackage{nicefrac}       
\usepackage{microtype}      
\usepackage{lipsum}
\usepackage{fancyhdr}       
\usepackage{graphicx}       
\graphicspath{{media/}}     

\usepackage{subfig}
\usepackage{amssymb} 
\usepackage{amsmath}

\usepackage{enumitem}

\title{A Multicast Scheme for Live Streaming Courses in Large-Scale, Geographically Dense Campus Networks
}

\author{
  Senxin Wu, Jinlong Hu, Ling Zhang* \\ 
  School of Computer Science and Engineering \\
  South China University of Technology \\
  Guangzhou, China\\
  \texttt{tenma@vip.qq.com, jlhu@scut.edu.cn, ling@scut.edu.cn} \\
}

\begin{document}
\maketitle

\begin{abstract}
Video courses have become a significant component of modern education. However, the increasing demand for live streaming video courses places considerable strain on the service capabilities of campus networks. The challenges associated with live streaming course videos in campus network environments exhibit distinct spatial distribution characteristics. The audience for specific video courses may be highly concentrated in certain areas, leading to a large number of users attempting to access the same live stream simultaneously. Utilizing a Content Delivery Network (CDN) to distribute videos in these campus scenarios creates substantial unicast pressure on edge CDN servers. This paper proposes a two-layer dynamic partitioning Recursive Bit String (RBS) virtual domain network layer multicast architecture specifically designed for large-scale, geographically dense multicast scenarios within campus networks. This approach reduces redundant multicast messages by approximately 10-30\% compared to the two-layer fixed partitioning method. Additionally, it establishes multicast source authentication capabilities based on Source Address Validation Improvement (SAVI) and facilitates secure multicast group key exchange using a concise exchange protocol within the WebRTC framework. In the next-generation data plane of programmable software-defined networks, the RBS stateless multicast technology can be integrated with the unique characteristics of large-scale, geographically dense campus network scenarios to dynamically and efficiently extend multicast coverage to every dormitory.
\end{abstract}


\section{Introduction}
In recent years, advancements in Internet technology and the emergence of services such as wired 4K, wireless 2K, and augmented reality/virtual reality (AR/VR) have ushered the world into an era of "video everywhere". Video services have evolved from singular media applications to a diverse array of interactive video applications that depend on robust network capabilities. Video courses have become an essential component of modern education. However, the growing demand for large-scale high-definition, ultra-high-definition, and even 3D video course live streaming is challenging the service capabilities of campus networks.

Currently, commercial online teaching platforms often utilize video servers in conjunction with Content Delivery Networks (CDN) and unicast networks for video transmission. As the number of viewers increases, these servers must implement load balancing to effectively manage demand. Limited network bandwidth adversely affects the quality and real-time performance of video playback. In the future, an increasing number of courses will be conducted via live streaming, which will exacerbate the existing bottleneck issues related to network bandwidth. The target audience for campus video courses is typically concentrated within a few campuses, resulting in significant unicast pressure on edge CDN servers, while other CDN servers struggle to alleviate this bandwidth strain. Under the CDN video distribution architecture, streaming a 1080p (60 FPS) H.265 encoded video (with a bitrate of 5 to 12 Mbps) requires approximately 117 servers with 10 Gbps bandwidth to support around 3,333 groups of 30-person course live streams, or about 468 servers with 2.5 Gbps bandwidth.

\begin{figure}
\centering
\includegraphics[width=0.5\linewidth]{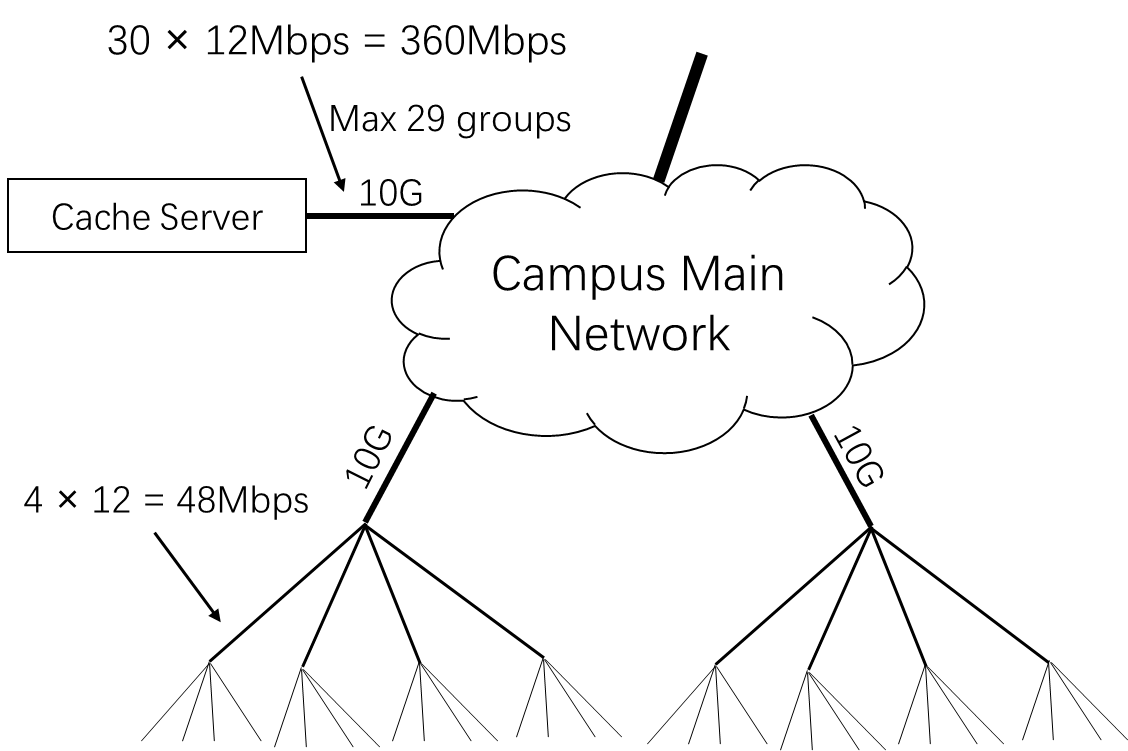}
\caption{\label{fig:CDN_campus_main_network}Class Video Live Streaming Based on CDN (1080P, 60FPS, H.265 Encoding).}
\end{figure}

Network layer multicast technology is an efficient one-to-many transmission method. By sending a data packet only once, multiple receivers can receive the same data, significantly reducing network bandwidth usage and server load. However, implementing large-scale multicasting still presents several challenges, including complex and scalable network topologies, the selection of appropriate multicast protocols, device compatibility issues, and security concerns.

The challenges of live video streaming for courses in campus network environments are characterized by their unique features, including large scale and dense spatial distribution. Different types of audience groups for video courses exhibit varying spatial distribution characteristics. Certain specific audience groups may demonstrate a particularly high density in their spatial distribution, such as:

\begin{itemize}

\item The audience groups for the video course will be concentrated on 1 $\sim$ 2 floors of 1 $\sim$ 2 dormitory buildings;

\item The video course audience groups, organized by colleges, will be concentrated in several nearby dormitory buildings;

\item The video course will target audience groups based on grade levels, covering nearly all dormitory buildings and the majority of dormitories within those buildings;

\item The audience groups for video courses based on university lectures consist of a large number of individuals, with their spatial distribution being both scattered and dense.

\end{itemize}

In this paper, we focus on enhancing live streaming services within campus networks and propose an IPv6+ multicast architecture designed for large-scale, geographically dense multicast scenarios. This scheme enables elastic scaling of IPv6 video multicasting within the campus network without requiring modifications to the client protocol stack. Additionally, it ensures the authenticity of multicast members and content, as well as the secure exchange of group keys.

\section{Related Works}

\subsection{Network Layer Multicast Technology}

Traditional multicast typically refers to stateful network layer multicast, which utilizes the Multicast Listener Discovery (MLD) protocol \cite{rfc3810} in conjunction with the Protocol Independent Multicast (PIM) protocol \cite{rfc3973} \cite{rfc2117} for network layer multicast in IPv6 networks. MLD is responsible for monitoring terminals that attempt to participate in multicast within the subnet at the edge routers, while the PIM protocol facilitates the exchange of multicast states between multicast routers to construct multicast distribution trees. Traditional multicast necessitates the storage of the state of each multicast group at every router along the multicast path, and the limited hardware resources of routers impose restrictions on the maximum number of multicast groups permitted within an autonomous network. The PIM protocol constructs multicast distribution trees through methods such as flooding, joining, and pruning; however, the time required to establish and migrate these multicast distribution trees is considerable.

SDN Multicast \cite{qi-bitar-intarea-sdn-multicast-overlay-01} leverages Software-Defined Networking (SDN) controllers to issue multicast flow table rules to each SDN router along the multicast path. This approach eliminates the high time costs associated with PIM methods, such as flooding, which are traditionally used to construct multicast distribution trees. Additionally, SDN controllers possess a global view of the network, enabling them to generate optimal multicast distribution trees based on the locations of multicast participants and the overall network topology. They can also deploy new multicast distribution trees immediately upon detecting congestion, thereby preventing potential bottlenecks. While SDN multicast enhances the efficiency and scalability of multicast distribution tree construction, it does not specify how to route multicast packets on the routers nor does it alleviate the pressure caused by the number of multicast groups on these devices.

XCast \cite{rfc5058} is a stateless multicast technology, which means that no intermediate routers are required to store any multicast state information. XCast incorporates the destination IP addresses of multiple multicast members into the IP packet header, allowing routers to forward packets based on several destination addresses without the need to maintain any multicast state information. Instead, they only require unicast routing information to facilitate the multicast routing function. However, embedding multiple multicast destination addresses in the IP packet header significantly limits the scalability of multicast due to the maximum transmission unit (MTU) size of the packet.

COXCast employs the Chinese Remainder Theorem \cite{pei1996chinese} to route multicast packets. The core concept involves assigning each router within the autonomous network a unique prime number \( K_i \), which abstracts the forwarding interfaces of the multicast distribution tree at each router into binary bit strings \cite{lacan2020xor}. Through complex mathematical functions, COXCast comprehensively calculates the corresponding primes and binary bit strings for all routers along the multicast path, ultimately deriving the encoding for the multicast group. Routers only need to compute the remainder of this multicast encoding by \( K_i \) to obtain the corresponding forwarding interface. PolKA \cite{dominicini2020polka} extends COXCast to the second-order Galois field, enabling multicast routing through binary remainders and enhancing multicast routing efficiency.

BIER (Bit-Indexed Explicit Replication) \cite{rfc8279} is a stateless multicast routing technology that indicates whether a destination participates in the current multicast group by marking a '1' in the corresponding position of a bit string. This bit string is then attached to the multicast packet to facilitate routing. Intermediate routers only need to determine how to copy and forward multicast packets based on the bit string and the Bit-Indexed Forwarding Table (BIFT). The multicast state information stored in the BIFT is not tied to specific multicast groups; rather, it pertains solely to the topology of the multicast network. Consequently, the number of multicast states at intermediate routers is independent of the number of multicast groups. However, BIER maps each multicast destination to a unique bit position, which means that the length of the bit string corresponds to the number of devices capable of participating in multicast within the network.

Carrier Grade Minimalist Multicast (CGM2) \cite{eckert-bier-cgm2-rbs-01}, also known as Recursive Bit String Structure (RBS) \cite{eckert-bier-rbs-00}, is a stateless multicast technology that employs recursive bit strings to encode multicast distribution trees. RBS utilizes bit strings to represent the forwarding ports of multicast packets on a single router, beginning from the root node of the multicast distribution tree and recursively encoding each subtree. Routers interpret the recursive bit string and route multicast packets based on the bit string associated with the current router. The length of the RBS is influenced by the scale of the multicast and the number of router ports that the multicast traverses; however, it is not affected by the overall scale of the multicast network.

Although BIER can be applied in large-scale multicast networks through hierarchical partitioning, conducting multicast across multiple subdomains results in significant redundant traffic within the backbone domain. Additionally, even small-scale multicast can produce considerable redundant traffic in the backbone domain; this issue is not related to the scale of the multicast itself but rather to the number of BIER subdomains involved. RBS addresses this problem by generating redundant traffic only during large-scale and sparse multicast operations. Furthermore, the length of RBS is influenced by the geographical distribution of multicast members, and small-scale yet sparse multicast distribution trees can lead to a sharp increase in the length of RBS.

RBS specifically defines the expression of multicast distribution trees and the routing of multicast packets. Notably, BIER can utilize RBS instead of flat bit strings, referred to as RBS-BIER \cite{eckert-bier-rbs-00}. Similarly, RBS can also be applied to MSR6 \cite{cheng-msr6-design-consideration-00} \cite{lx-msr6-rgb-segment-05} \cite{geng-msr6-rlb-segment-02}, i.e., \cite{xu-msr6-rbs-01} \cite{eckert-msr6-rbs-01}.

The primary challenge of network layer multicast lies in the conflict between the number of states that can be represented by the multicast header and the limited hardware resources of network devices. Implementing large-scale multicast in extensive networks necessitates a significant number of states to accurately describe the expansive multicast distribution tree. Stateless multicast requires encoding this extensive distribution tree using a substantial number of bits in the multicast header, whereas stateful multicast necessitates the prior recording of relevant state information for multicast groups within network devices.

\section{Problem Model and Research Boundaries}

Assuming the upper limit of the school's student population is 100,000, with 30 students per class, this means that 3,333 teachers are simultaneously conducting live video courses for 100,000 students. The scale of each group's live course can be categorized into the types listed in Table 1. Additionally, it is assumed that the video bandwidth is set at 1080P (60 FPS, 12 Mbps) and 4K (60 FPS, 62 Mbps).

\begin{table}[h]
\centering
\caption{\label{tab:course_classification}Classification of Campus Video Courses.}
\begin{tabular}{lcc}
\hline
Course Type & Number of Participants & Percentage \\
\hline
Small Class & 30 people & 50\% \\
Professional Course & 30$\sim$90 people & 20\% \\
Public Course & 60$\sim$150 people & 20\% \\
Mini-lecture & 200 people & 5\% \\
Large Lecture & 1000 people & 4\% \\
Mega-lecture & 10000 people & 1\% \\
\hline
\end{tabular}
\end{table}

The challenge of broadcasting live video courses over campus networks can be characterized as a large-scale, geographically dense multicast scenario, which encompasses four main features: 1) A significant number of network participants within a relatively straightforward topology; 2) Multicast members may be densely distributed in space; 3) There is a substantial number of devices anticipating participation in the multicast network; 4) Terminal devices seek to join the multicast network with minimal complexity.

The design goal of the large-scale, geographically dense multicast architecture for campus networks is to achieve elastic-scale IPv6 video multicast without modifying the client protocol stack. Additionally, it aims to ensure the authenticity of multicast members and content, as well as the secure exchange of group keys. This design goal is accomplished by dividing the architecture into three modules: IPv6+ network layer multicast, WebRTC, and a multicast authentication framework. This paper focuses on two key issues: 1) In the geographically dense multicast scenario of campus networks, the RBS multicast technology based on Software-Defined Networking (SDN) is selected, and RBS hierarchical partitioning is implemented for this scenario; 2) The extension of the Diffie-Hellman (DH) exchange for multicast is proposed to facilitate the secure exchange of group keys.

\section{Module Design}
\subsection{Overall System Architecture}

To address the large-scale, geographically dense multicast scenarios in campus networks, we have designed a network layer multicast architecture based on SDN with dynamic partitioning of RBS virtual domains over IPv6. Simultaneously, we have extracted business-irrelevant functions of video live streaming into a Video Live Streaming Management System (VLSMS), which serves as a public infrastructure for various upper-layer business applications. The VLSMS interacts with the Network Management System (NMS) and, through the SDN controller, sends multicast flow table rules to the campus network's SDN switches to direct the routing of multicast packets. Users can seamlessly join the multicast via the WebRTC module on their terminal devices through a web browser. The VLSMS intercepts and manages the startup process of WebRTC, scheduling the corresponding multicast capabilities through the NMS. Finally, this paper presents a multicast authentication framework that includes: 1) Issuing legal multicast group rules (quadruples composed of source address, destination multicast address, source port, and destination multicast port) to SAVI agents to support multicast source authentication; 2) Extending the Diffie-Hellman (DH) exchange for multicast and employing cubic spline interpolation to facilitate the secure exchange of group keys.

\begin{figure}[h]
\centering
\includegraphics[width=1\linewidth]{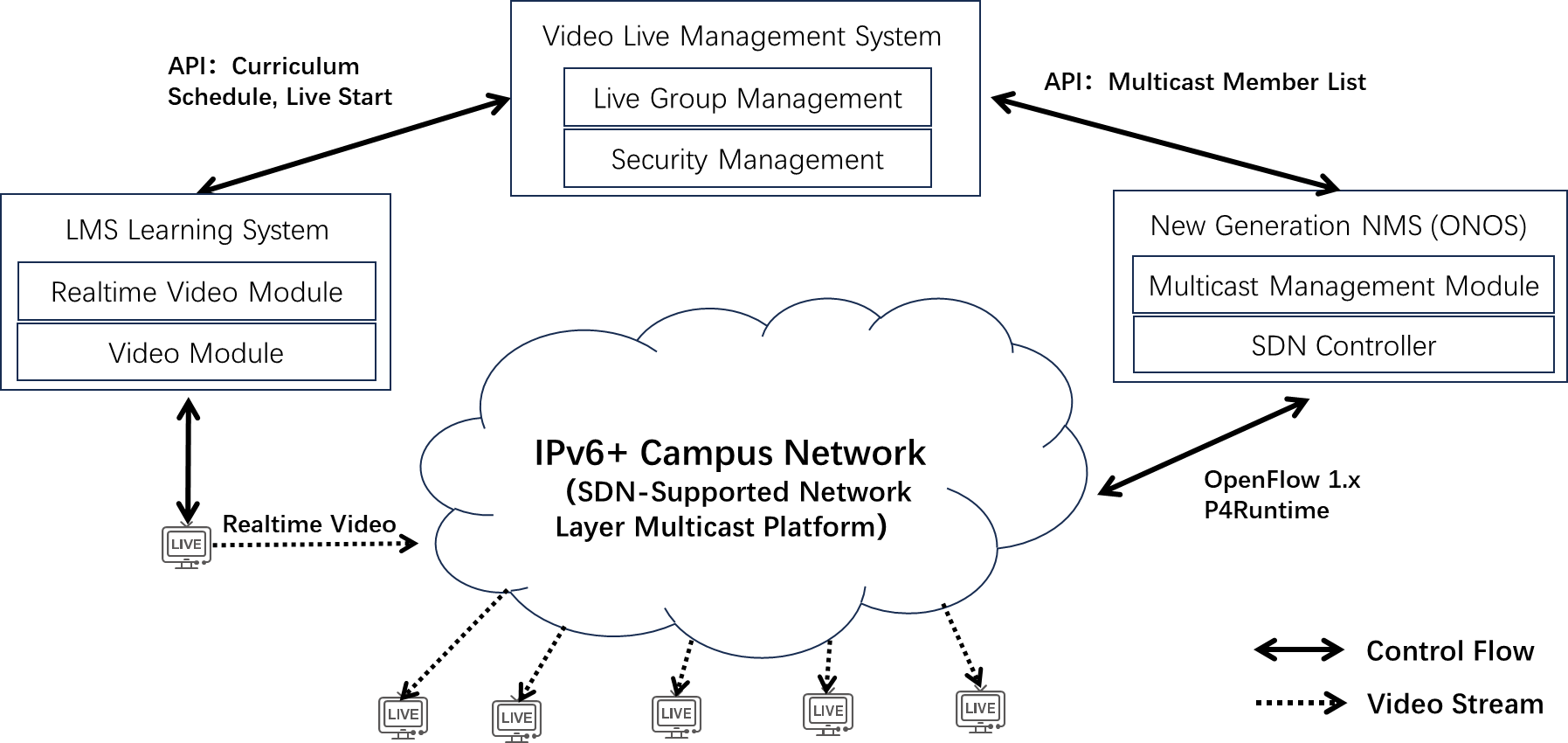}
\caption{\label{fig:system_arch}Campus Network Large-scale Geographically Dense Multicast System Architecture.}
\end{figure}

\subsection{Network Layer Multicast Architecture}

Since not all network devices can support multicast, the campus network can be broadly categorized into multicast and unicast layers. A wider multicast range results in fewer redundant packets within the unicast layer, leading to higher efficiency in video live streaming. The design objective of the campus network's multicast architecture should be to ensure that the multicast range effectively reaches every dormitory. Once the multicast range extends to each dormitory, it indicates that the multicast network must accommodate the multicast demands of 100,000 network devices. Therefore, whether utilizing SDN multicast, BIER multicast, or RBS multicast, each approach has its limitations in large-scale, geographically dense multicast scenarios. SDN multicast necessitates the issuance of multicast flow table rules to all routers along the multicast path; BIER multicast requires a large bit string to uniquely identify all devices within the multicast network; and RBS multicast also demands a large bit string to encode the complete multicast path.

Using a hierarchical domain architecture can segment an extensive campus multicast network into several subdomains. By employing SDN technology to implement inter-domain multicast conversion flow tables between these subdomains, redundant multicast traffic within each domain can be significantly reduced. Consequently, only a limited number of inter-domain routers are required to issue flow table rules. Utilizing RBS multicast technology within subdomains can effectively address the challenge of rapid expansion in the multicast encoding length embedded in IPv6 packets, which is often exacerbated by an increase in the number of network devices. Instead, the encoding length is influenced by the current scale of multicast group members and the multicast path, rather than being directly related to the total number of network devices within the subdomain. In large-scale, geographically dense multicast scenarios within the same subdomain, RBS multicast is expected to generate less redundant traffic compared to BIER multicast (refer to CGM2\cite{eckert-bier-cgm2-rbs-01}).

To achieve this, this paper divides the campus network into a double-layer multi-RBS subdomain structure, ensuring that every dormitory on campus is covered within the multicast range. Additionally, a dynamic partitioning RBS virtual domain algorithm is proposed to minimize redundant packets, reduce flow table operations, and expand the multicast range within the network.

\begin{figure}[h]
\centering
\includegraphics[width=0.5\linewidth]{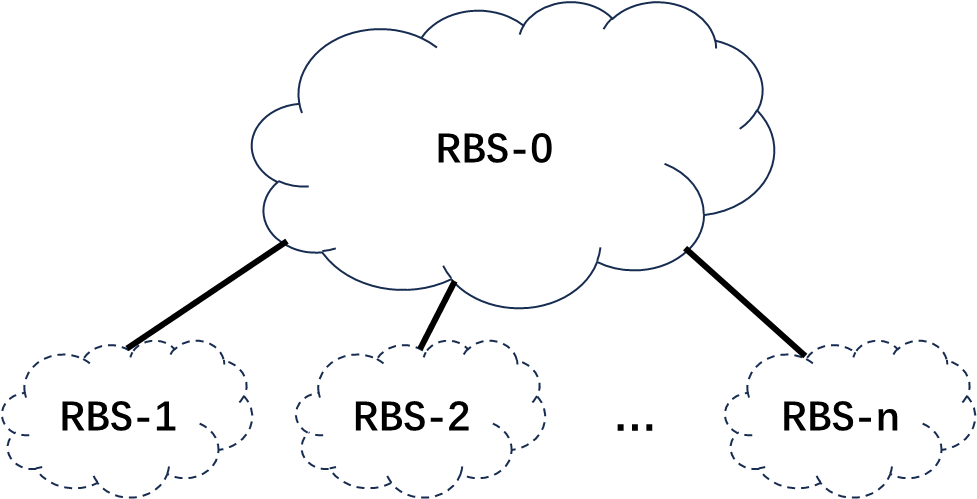}
\caption{\label{fig:system_arch}Network Layer Multicast Architecture.}
\end{figure}

The concept of dynamic partitioning of virtual domains refers to a network device that is not permanently assigned to a specific domain. Instead, it dynamically adjusts its assigned domain based on the current status of multicast groups.

Represent the campus network topology as a graph, denoted as \( G = \{V, E\} \), where \( V = \{v_1, v_2, \ldots, v_m\} \) and \( m \) represents the number of routing and switching nodes in the network. The set \( E = \{e_1, e_2, \ldots, e_k\} \) contains the links in the network, with \( k \) indicating the total number of links. Each link \( e_i \) is defined as \( \{linkId, srcNode, dstNode, bandWidth\} \). Let the multicast distribution tree be represented as \( TP = \{TV, TE\} \), where \( TV \) includes all nodes in the graph from the multicast source to the multicast receiver (\( TV \subseteq V \)), and \( TE \) comprises the links associated with the nodes along the path (\( TE \subseteq E \)). Additionally, let \( edgeTV \) denote the leaf nodes of the multicast distribution tree, which are the multicast receivers. 

The optimization goal of the dynamic partitioning RBS virtual domain algorithm is to minimize the number of redundant multicast packets based on the distribution of multicast members at each instance. This can be expressed mathematically as:
\begin{equation}
\underset{\forall RL_i \leq maxRBSLength}{\text{argmin}} \ j,\space RL_i = \text{Length}(RBSEncode(groupEdgeTV_i)), \quad i = 1, 2, \ldots, j.
\end{equation}

Randomly select leaf nodes from the multicast distribution tree to form teams (denoted as \( groupEdgeTV_i \)), aiming to minimize the number of teams \( j \) under the condition:
\begin{equation}
 \underset{\forall RL_i \leq maxRBSLength}{\text{argmin}} \ j, \quad maxRBSLength \in \{256 \text{ bits}, 512 \text{ bits}, 1024 \text{ bits}\}. 
\end{equation}

The basic steps of the dynamic partitioning RBS virtual domain algorithm: 
\begin{enumerate}[label=(\arabic*)]
\item \label{item:calculate_tree} Calculate the multicast distribution tree based on the distribution of multicast members;

\item \label{item:calculate_distance} Calculate the distance between multicast members;

\item \label{item:form_virtual_domains} Multicast members establish virtual domains nearby;

\item \label{item:select_domain} Select the virtual domain with the highest number of multicast members based on the specified RBS encoding length;

\item \label{item:reform_domains} The remaining multicast members of non-selected virtual domains re-form domains nearby. The distance function can be defined as the number of router hops between two points. 
\end{enumerate}

When the Floyd-Warshall algorithm is employed to calculate the distances between all nodes in a graph, its time complexity is \(O(m^3)\). The time complexity for sorting nearby multicast members is \(O(m^2 \log m)\). Therefore, the overall time complexity of the algorithm is \(O(m^3 + m^2 \log m + j \cdot m^2)\). Under the condition of maintaining the ascending order in steps \ref{item:form_virtual_domains} and \ref{item:select_domain}, and assuming no significant changes in the distances between multicast members, binary search can optimize the algorithm's time complexity to \(O(m^2 \log m + j \cdot m \log m)\).

The fundamental proof of the dynamic partitioning RBS virtual domain algorithm is as follows: 1) The RBS encodes all links traversed by the multicast distribution tree, marking links that need to copy packets as 1 and others as 0. Starting from a specific leaf node (a multicast member), nearby domains can identify the virtual domain with the highest number of multicast members originating from that node. 2) By iteratively selecting the virtual domain with the most multicast members, the total number of domains, denoted as j, will eventually be minimized. This is achieved by excluding several multicast members from the initially identified virtual domain with the highest count, thereby reallocating these members to other virtual domains. In this scenario, the optimal outcome is for these multicast members to belong to their directly adjacent virtual domain. It is important to note that transferring multicast members to adjacent virtual domains cannot decrease the total number of virtual domains, as these adjacent domains are already at capacity. Consequently, the number of multicast members in adjacent virtual domains remains unchanged after the adjustment of multicast members.

The basic proof of the dynamic partitioning RBS virtual domain algorithm: 1) Since RBS will encode all links passed by the multicast distribution tree (links that need to copy packets are marked as 1, and other links are marked as 0), starting from a certain leaf node (multicast member) must be nearby domains can get the virtual domain with the most multicast members starting from that node; 2) Iteratively selecting the virtual domain with the most multicast members will eventually make the number of domains j minimum: Exclude several multicast members from the first obtained virtual domain with the most multicast members, so that these multicast members belong to other virtual domains, at this time according to 1) the best case is to belong to its directly adjacent virtual domain, and the operation of belonging to adjacent virtual domains cannot reduce the number of virtual domains, because adjacent virtual domains are already full, that is, the same number of adjacent virtual domains do not increase the number of multicast members contained after adjusting multicast members.

\subsection{Multicast Group Key Exchange Algorithm}

Extension of DH exchange for multicast scenarios: 1) The key negotiator generates a confidential random number \( a \) and shared parameters \( g \), \( p \), and \( A \), where \( A = g^a \mod p \); 2) Each multicast member \( i \) generates a confidential random number \( x_i \) and shared parameter \( B_i = g^{x_i} \mod p \); 3) The key negotiator shares \( B_i \) with the multicast members and calculates \( K_i = A^{x_i} \mod p = B_i^a \mod p \); 4) The key negotiator randomly generates a multicast group shared key \( y \), thereby obtaining a set of point pairs \( (K_i, y) \); 5) The key negotiator employs cubic spline interpolation to fit the key function \( y = S(K_i) \) and informs each multicast member of the function \( S \). 

\begin{figure}[h]
    \centering
    \subfloat[Diffie-Hellman Key Exchange Multicast Extension.\label{fig:DH_Exchange_Multicast_Extension}]{%
        \includegraphics[width=0.45\textwidth]{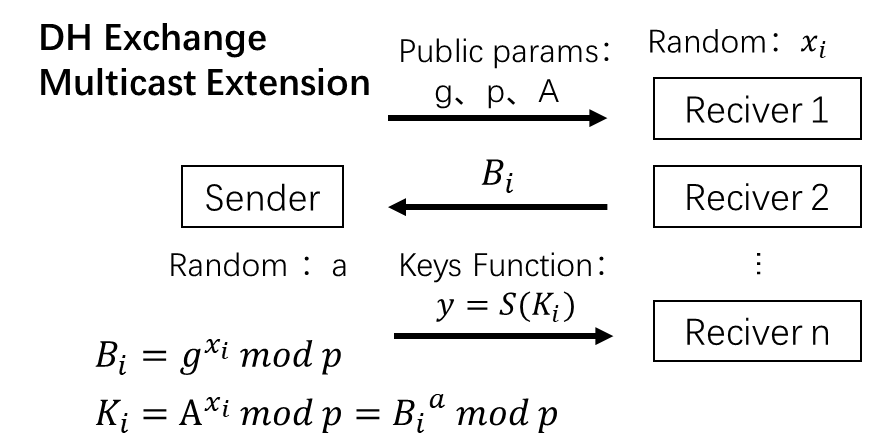}}
    \hfill
    \subfloat[Cubic Spline Interpolation Method.\label{fig:spline_interpolation}]{%
        \includegraphics[width=0.45\textwidth]{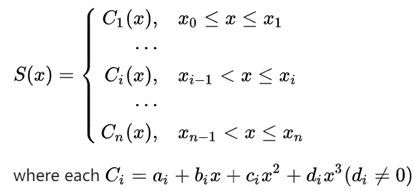}}
        \caption{\label{fig:DH exchange}Multicast Group Key Exchange Algorithm.}
\end{figure}

Since cubic spline interpolation employs piecewise functions to fit pairs of points, the function \( S \) contains only one multicast member point pair \((K_i, y)\) in each interval. This indicates that there is no significant correlation between different multicast members, meaning that the joining or leaving of multicast members will not result in the leakage of the shared key. To enhance the security of the multicast content, it is essential to expand the range of \( y \) and \( K_i \) as much as possible and to regularly update the shared key. Additionally, it is crucial to replace the shared key whenever multicast members exit the multicast group.

\begin{figure}[h]
\centering
\includegraphics[width=0.75\linewidth]{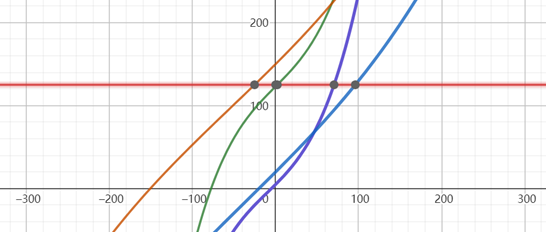}
\caption{\label{fig:spline_interploation_function}Key Function $y = S(K_i)$.}
\end{figure}

\section{Simulation and Result Analysis}

Simulations of the dynamic partitioning RBS virtual domain algorithm were conducted within the large-scale, geographically dense scenario of the campus network. The topology of the campus network is divided into a core network and an edge network. The core network comprises approximately 100 core routers, while the edge network includes core edge routers, secondary edge routers, and user access routers. Core routers serve as the backbone of the campus network, core edge routers facilitate the connection between the core network and the edge network, user access routers link the edge network to user devices, and edge routers connect core edge routers to user access routers.

By executing the dynamic partitioning RBS virtual domain algorithm and the fixed partitioning RBS domain algorithm on randomly generated campus network topologies, we compare the performance differences between the two algorithms under average conditions. The following are the rules for generating random campus network topologies:

\begin{itemize}

\item Specify the number of core routers, and the probability of a link existing between any two core routers is 10\%;

\item Specify the number of edge routers, and the probability of a link existing between any two edge routers is 10\%;

\item Core edge routers randomly connect to a designated number of core routers;

\item The number of links between core edge routers and core routers follows a log-normal distribution, denoted as \(X \sim f(x; 2, 1.5)\).

\begin{equation}
f(x; \mu, \sigma) = \frac{1}{x \sigma \sqrt{2\pi}} e^{-\frac{(\ln x - \mu)^2}{2\sigma^2}}
\end{equation}

\item Specify the number of user access routers, and user access routers randomly linked to a secondary edge router;

\item Specify the number of user devices, and user devices are randomly linked to a user access router.

\end{itemize}

The simulation network consists of 60 core routers, 128 core edge routers, 12 secondary edge routers, 128 user access routers, and 512 user devices. The core routers and core edge routers form the backbone of the simulation network. The 12 secondary edge routers are utilized to simulate user access within 12 regions, representing a large-scale, geographically dense scenario. This setup focuses solely on simulating the local topology of the campus network.

The multicast density within a specified range is categorized into ten levels, from sparse to dense (0–100

The maximum packet length for IPv6 packets is generally 1500 bytes. Therefore, the recommended values for RBS encoding lengths in the IPv6 extension header are 256 bits, 512 bits, and 1024 bits. The fixed partitioning RBS domain algorithm assigns user devices in a specific area to a designated core edge router. When the RBS encoding space is relatively ample and the multicast density is high, a single multicast packet may suffice to include all multicast members within a fixed domain. Due to the large number of multicast members in this domain, the utilization rate of the RBS encoding space is relatively high, resulting in a minimal difference in spatial utilization between the two algorithms. However, when the RBS encoding space is still sufficient but the multicast density is low, the multicast distribution paths in different domains do not fully utilize the RBS encoding space. In this scenario, the dynamic partitioning RBS virtual domain algorithm effectively leverages these available encoding spaces, thereby achieving higher multicast efficiency. Smaller RBS encoding spaces (256 bits) further amplify the differences in spatial utilization between the two algorithms.

Within a domain, if the multicast distribution tree does not fully utilize the RBS encoding space, the dynamic domain method significantly enhances the utilization rate of the RBS encoding space compared to the fixed domain method. When the encoding space is relatively abundant, the advantages of the dynamic domain method become more pronounced, particularly when the same scale of multicast groups is sparsely distributed. Conversely, when multicast groups are densely distributed, the dynamic domain method aligns with the fixed domain method. In scenarios where the available encoding space is limited, the dynamic domain method also aligns with the fixed domain method. It is essential to specify the scale of multicast members and adjust the multicast density within domains. Specifically, when domain density is high, multicast members will be concentrated in a few domains (dense); when domain density is low, multicast members will be spread across multiple domains (sparse).

The dynamic partitioning RBS virtual domain algorithm and the fixed partitioning RBS domain algorithm require fewer flow table entries (the number of routers that need to be managed).

\section{Conclusion}

Video courses are playing an increasingly important role in modern education, which poses significant challenges to the service capabilities of campus networks, particularly in light of the rising demand for live streaming. The issue of live course video streaming within campus networks exhibits unique spatial distribution characteristics, as audiences for specific courses may be geographically concentrated, with many users expecting to access the same live stream simultaneously. In such an environment, utilizing a CDN to distribute course videos can create substantial unicast pressure on edge servers. To address this challenge, this paper proposes an innovative double-layer dynamic partitioning RBS virtual domain network layer multicast architecture, optimized for large-scale, geographically dense multicast scenarios typical of campus networks. Compared to the double-layer fixed partitioning method, this approach can reduce redundant multicast packets by approximately 10\% to 30\%. Additionally, we have developed a multicast source authentication mechanism that integrates SAVI and a multicast group key secure exchange mechanism based on DH exchange via WebRTC. In the future, in the context of next-generation data plane programmable software-defined networks, we aim to combine RBS stateless multicast technology with the unique characteristics of campus networks to achieve efficient dynamic expansion of the multicast range, ensuring that every dormitory can enjoy a seamless live course streaming experience.

\bibliographystyle{unsrt}  
\bibliography{references}

\end{document}